\documentclass[12pt]{iopart}
\usepackage{graphicx}
\usepackage{bm}
\eqnobysec

\newcommand{\beq}{\begin{equation}}
\newcommand{\beqa}{\begin{eqnarray}}
\newcommand{\eeq}{\end{equation}}
\newcommand{\eeqa}{\end{eqnarray}}
\newcommand{\abs}[1]{\vert{#1}\vert}
\renewcommand{\d}{{\rm d}}
\renewcommand{\e}{{\rm e}}

\renewcommand{\es}{{\vphantom{M}}}
\newcommand{\f}{{\rm f}}
\newcommand{\frad}[2]{\displaystyle{\displaystyle#1\over\displaystyle#2}}
\newcommand{\ii}{{\rm i}}
\renewcommand{\max}{{\rm max}}
\newcommand{\mean}[1]{\langle#1\rangle}
\renewcommand{\phi}{\varphi}
\newcommand{\rr}{{\rm r}}
\renewcommand{\tr}{\mathop{\mathrm{tr}}\nolimits}
\newcommand{\ve}[1]{{\bm#1}}

\newcommand{\A}{{\ve A}}
\newcommand{\B}{{\ve B}}
\newcommand{\C}{{\ve C}}
\newcommand{\E}{{\cal E}}
\renewcommand{\H}{{\cal H}}

\begin{document}

\title
[Tight-binding spectra on spherical graphs I: the effect of a magnetic charge]
{Tight-binding electronic spectra on graphs with spherical topology I:
the effect of a magnetic charge}

\author{Y Avishai$^{1,2,3,4}$ and J M Luck$^3$}

\address{$^1$ Department of Physics and Ilse Katz Center for Nanotechnology,
Ben Gurion University, Beer Sheva 84105, Israel}

\address{$^2$ RTRA -- Triangle de la Physique, Les Algorithmes, 91190
Saint-Aubin, France}

\address{$^3$ Institut de Physique Th\'eorique\footnote{URA 2306 of CNRS},
CEA Saclay, 91191~Gif-sur-Yvette cedex, France}

\address{$^4$ Laboratoire de Physique des Solides\footnote{UMR 9502 of CNRS},
Universit\'e Paris-Sud, 91405 Orsay cedex, France}

\begin{abstract}
This is the first of two papers devoted to tight-binding electronic spectra
on graphs with the topology of the sphere.
In this work the one-electron spectrum is investigated
as a function of the radial magnetic field produced by a magnetic charge
sitting at the center of the sphere.
The latter is an integer multiple
of the quantized magnetic charge of the Dirac monopole,
that integer defining the gauge sector.
An analysis of
the spectrum is carried out for the five Platonic solids (tetrahedron, cube,
octahedron, dodecahedron and icosahedron), the C$_{60}$ fullerene,
and two families of polyhedra, the diamonds and the prisms.
Except for the fullerene, all the spectra are obtained in closed form.
They exhibit a rich pattern of degeneracies.
The total energy at half filling is also evaluated
in all the examples as a function of the magnetic charge.
\end{abstract}


\eads{\mailto{yshai@bgu.ac.il},\mailto{jean-marc.luck@cea.fr}}

\maketitle

\section{Introduction}
\label{intro}

Electronic properties of low-dimensional systems
have been the focus of intense research
for several decades~\cite{Imry,Giamarchi,Vignale}.
Electrons can be confined in various geometries,
such as a plane (two-dimensional electron gas),
a one-dimensional (quantum) wire,
or a quantum dot, which can be referred to as a zero-dimensional system.
In each one of these geometries,
new and spectacular physical phenomena have been exposed:
quantum Hall effect (in two dimensions), Luttinger liquid (in one dimension),
Coulomb blockade and Kondo effect (in zero dimension),
to mention just a few examples.

The present work is concerned with another realization
of a low-dimensional system which has so far received much less attention,
namely where electrons are confined to move on a compact surface.
The simplest class of such surfaces has the topology of the sphere.
The fullerene and its derivatives provide
the most natural candidates for such systems.
The main goal of the present work
is to investigate the single-particle energy spectrum
as a function of a radial magnetic field,
whereas the effect of spin-orbit interactions in the presence
of a radial electric field is the subject of the companion paper~\cite{II}.

Although the design of a pertinent experiment seems to be a formidable task,
there are several motivations for pursuing this topic on the theoretical front.
First, one encounters here a zero-dimensional system with
a topology different from the usual one of a quantum box, say.
It is surprising that, although the C$_{60}$ fullerene
was discovered more than two decades ago~\cite{Kroto},
the study of electronic properties of nanoscopic systems
with the topology of the sphere is still rather scarce.
It is indeed expected that topology and geometry
should play an important role in the physics of electrons
residing on closed surfaces.
The examples treated in this work
will demonstrate that the tight-binding spectra
of polyhedra exhibit a rich pattern of degeneracies,
at variance with the regularity
of the spectrum of the continuum Schr\"odinger equation
on the sphere in the presence of a magnetic charge~\cite{tamm}.
A similar scenario was already known to occur in the case of the plane,
when the simple pattern of Landau levels for an electron
subject to a perpendicular magnetic field is
compared with the strikingly rich Hofstadter butterfly~\cite{Butter} resulting
when the same problem is treated on the square lattice,
within the tight-binding scheme.
Among more direct motivations, let us recall
that it has been suggested that fullerenes, i.e., carbon molecules
with spherical topology, can be described in an effective way
by the Dirac equation in the continuum
in the presence of the magnetic field of an effective quantized monopole
of charge one-half~\cite{Guinea}, in the absence of any physical magnetic field.
Moreover, it was shown recently that magnetic monopoles may emerge
in a class of exotic magnets known collectively as spin ice~\cite{Sondhi}.
Third, the idea of considering electrons confined to
a spherical shell in the continuum and subject to a
magnetic field emerging from a magnetic monopole proved to be useful in the
study of the quantum Hall effect~\cite{Haldane,AHK}, as it avoids the
boundary problem.

Having laid down our motivations,
let us now move on and detail our work.
The objective is to investigate the dependence of the tight-binding
energy spectrum on the value of a radial magnetic field,
for a wide range of examples of graphs with spherical topology.
More concretely, the electron lives on the vertices (sites)
of a polyhedron drawn on the unit sphere and executes nearest-neighbor hopping.
The energy levels are examined as a function
of a radial isotropic magnetic field.
The crucial difference between
the present study and that of the Hofstadter problem~\cite{Butter} is that here,
the only way to create a radial magnetic field is by placing a magnetic charge
at the origin, i.e., at the center of the sphere.
As is well known, such a magnetic charge must be
quantized~\cite{Dirac}, i.e., be equal to the product
of the elementary magnetic charge $g=\hbar c/(2e)$
of the famous Dirac monopole by an integer~$n$.
The total magnetic flux coming out through the sphere is $n\Phi_0$,
where $\Phi_0=4\pi g=hc/e$ is the flux quantum.
The (positive or negative) integer $n$,
hereafter referred to as the magnetic charge,
determines the gauge sector of the problem.
Within the tight-binding model,
the magnetic field enters the problem through
U(1) phase factors living on the oriented links.
The product of the phase factors living on the anticlockwise
oriented links around a face of the polyhedron equals
$\exp(2\pi\ii\phi/\Phi_0)$,
where $\phi$ is the outgoing magnetic flux through that face.
This discrete formalism allows one to avoid the singularities
met in the continuum description of monopoles~\cite{Yang}.
For a given graph drawn on the unit sphere, e.g.~a regular polyhedron,
the energy spectrum and all the observables depend solely on
the magnetic charge~$n$.
Although it might be tempting to regard the value of $n$ as
representing the strength of the magnetic field,
this viewpoint might be somewhat misleading.
Recall that the Dirac quantization condition
implies that there is no continuous deformation
of one gauge sector into another.
Furthermore, the results of this work will demonstrate
that there is in general no simple relationship between the value of $n$
and the observed pattern of level degeneracies.
As already mentioned, the present situation is in strong contrast with
its continuum analogue, namely that described by the Schr\"odinger equation
on the sphere in the presence of a magnetic monopole,
investigated in the pioneering work of Tamm~\cite{tamm},
whose spectrum has a very regular dependence on the magnetic charge $n$.
The energy levels indeed read
\beq
E_{n,\ell}=\frac{\hbar^2}{2MR^2}
\left(\ell(\ell+1)+\abs{n}\left(\ell+\frac{1}{2}\right)\right),
\label{etamm}
\eeq
with $M$ being the mass of the particle,
$R$ the sphere radius, and $\ell=0,1,\dots$ the angular quantum number.
The corresponding multiplicities are
\beq
m_{n,\ell}=2\ell+\abs{n}+1.
\label{mtamm}
\eeq

The setup of the present paper is the following.
The general concepts and notations are introduced in Section~\ref{general}.
In Section~\ref{plato} the five regular polyhedra
or Platonic solids (tetrahedron, cube, octahedron, dodecahedron, icosahedron)
are considered.
The $F$ faces of these polyhedra are equivalent,
so that the problem is periodic in the integer $n$, with period $F$.
Section~\ref{sixty} is devoted to the investigation of the C$_{60}$ fullerene,
modeled as a symmetric truncated icosahedron
(where all links have the same length)
made of 12 pentagons (with solid angle $\Omega_5$)
and 20 hexagons (with solid angle $\Omega_6$).
The ratio $\Omega_5/\Omega_6$ is irrational,
so that the spectrum depends on the magnetic charge~$n$
in a quasiperiodic fashion.
Finally in Section~\ref{last} two families of polyhedra are considered:
the $N$-prism (rectangular prism with a regular $N$-gonal basis)
and its dual the $N$-diamond.
The spectra of these two examples are explicitly derived for all $N\ge3$.
Section~\ref{discussion} contains a discussion.
Miscellaneous results on the geometrical characteristics of the C$_{60}$
fullerene are presented in a self-consistent fashion in the Appendix.

\section{Generalities}
\label{general}

Throughout the following, the objects to be considered are polyhedral graphs
drawn on the unit sphere.
At the origin (center of the sphere) sits a
quantized magnetic charge equal to $n$ times the magnetic
charge $g$ of a Dirac monopole,
so that the total outgoing magnetic flux through the sphere is $n$
times the flux quantum $\Phi_0$.

Let us denote by $V$ the number of vertices (sites),
by $L$ the number of links (bonds) and by $F$ the number of faces
of the polyhedron.
In the case of the spherical topology, the Euler relation reads
(see e.g.~\cite{wilson})
\beq
V-L+F=2.
\label{euler}
\eeq

The tight-binding Hamiltonian of the system is
\beq
\hat\H=\sum_{<ij>}\left(a^\dag_iU_{ij}a_j+{\rm h.c.}\right),
\label{ham}
\eeq
where the sum runs over the $L$ oriented links $<\!ij\!>$ of the polyhedron,
and the~$U_{ij}$ are phase factors (elements of the gauge group U(1))
living on these links.
The Hamiltonian~$\hat\H$ is complex, i.e., not invariant under time reversal.
One has
\beq
U_{ij}=U_{ji}^{-1}=U_{ji}^\star,
\label{uustar}
\eeq
where the star denotes complex conjugation.
The product of the phase factors
around each face is related to the outgoing magnetic flux $\phi$
through the face~as
\beq
U_{ij}U_{jk}\dots U_{mi}=\exp(2\pi\ii\phi/\Phi_0).
\label{flux}
\eeq

The Hamiltonian $\hat\H$ is thus represented by a $V\times V$
complex Hermitian matrix $\H$ such that $\H_{ij}=U_{ij}$.
The equation for the energy eigenvalues $E_a$
and corresponding eigenfunctions $\psi_{i,a}$ reads
\beq
E_a\psi_{i,a}=\sum_{j(i)}U_{ij}\psi_{j,a},
\label{eigen}
\eeq
with $a=1,\dots,V$, whereas $j(i)$ runs over the neighbors of $i$.
The energy spectrum obeys the following sum rules
\beq
\sum_aE_a=0,\quad
\sum_aE_a^2=2L,
\label{sumrules}
\eeq
where the energy eigenvalues are counted with their multiplicities.
The first sum indeed equals $\sum_aE_a=\tr\H=\sum_i\H_{ii}=0$.
This zero sum rule holds whenever the Hamiltonian
only has non-diagonal matrix elements.
The second sum equals $\sum_aE_a^2=\tr\H^2=\sum_{ij}\abs{\H_{ij}}^2=2L$,
as each link gives two contributions equal to unity.

The main step in the explicit construction of the Hamiltonian matrix $\H$
consists in finding a configuration of the gauge field,
i.e., a set of phase factors $U_{ij}$,
so that~(\ref{flux}) holds for all the $F$ faces.
The $L$ phase factors $U_{ij}$ are therefore constrained by~$F$ conditions.
It will be clear from the example of the tetrahedron
that only $F-1$ of them are independent.
Using~(\ref{euler}), one is then left with $L-(F-1)=V-1$
degrees of freedom in the choice of a gauge field.
It proves convenient to fix these degrees of freedom
by setting all the phase factors of the links of a spanning tree
to their trivial values $U_{ij}=1$,
along a line of thought dating back to~\cite{stree}.
It should be recalled that a spanning tree is a tree,
i.e., a graph with no loop,
that spans the graph, i.e., passes through every vertex of the graph
(see e.g.~\cite{wilson}).
The choice of a gauge thus boils down to the discrete choice of a spanning
tree.
The number of non-trivial phase factors is equal to its minimal value $F-1$.

As already announced,
we shall successively investigate in detail the following examples:
the five regular polyhedra or Platonic solids,
the C$_{60}$ fullerene (modeled as a symmetric truncated icosahedron)
in Section~\ref{sixty}, and finally two families of polyhedra,
the diamonds and the prisms, in Section~\ref{last}.

\section{Platonic solids}
\label{plato}

There are only five regular polyhedra
in three dimensions, the Platonic solids.
This fact known since Antiquity can be easily recovered as follows.
Let $p$ be the coordination number of a vertex
and $q$ be the number of sides of a face.
Evidently, the equality $2L=pV=qF$ holds.
The Euler relation~(\ref{euler}) then implies
\beq
\frac{1}{p}+\frac{1}{q}=\frac{1}{2}+\frac{1}{L}.
\label{eulerfrac}
\eeq
The five solutions to~(\ref{eulerfrac}) with $p\ge3$ and $q\ge3$
correspond to the Platonic solids, as recalled in Table~\ref{tabplato}.
The tetrahedron is its own dual, whereas (cube, octahedron)
and (dodecahedron, icosahedron) form dual pairs
(see e.g.~\cite{wilson}).

\begin{table}[!ht]
\begin{center}
\begin{tabular}{|l||c|c||c|c|c|}
\hline
polyhedron&$p$&$q$&$V$&$L$&$F$\\
\hline
tetrahedron&3&3&4&6&4\\
cube&3&4&8&12&6\\
octahedron&4&3&6&12&8\\
dodecahedron&3&5&20&30&12\\
icosahedron&5&3&12&30&20\\
\hline
\end{tabular}
\end{center}
\caption{The five regular polyhedra or Platonic solids,
with their coordination numbers $p$, numbers $q$ of sides of a face,
and total numbers $V$ of vertices, $L$ of links and $F$ of faces.}
\label{tabplato}
\end{table}

The $F$ faces of a regular polyhedron are equivalent.
In particular, they have equal solid angles.
The magnetic flux through each phase is therefore $\phi=n\Phi_0/F$,
so that the right-hand side of~(\ref{flux}) reads
\beq
\omega=\exp(2\pi\ii n/F)
\label{omegadef}
\eeq
and obeys
\beq
\omega^F=1.
\label{omegaf}
\eeq
This property ensures that the problem is periodic in the integer $n$,
with period $F$.

\subsection{The tetrahedron}

\begin{figure}[!ht]
\begin{center}
\includegraphics[angle=90,width=.25\textwidth]{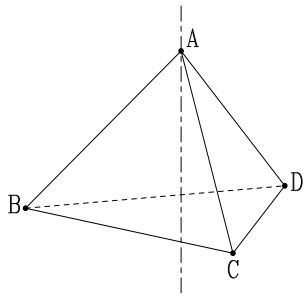}
\caption{
The tetrahedron.
The dashed-dotted line shows the threefold axis used to unwrap the structure.}
\label{tetra}
\end{center}
\end{figure}

The tetrahedron, shown in Figure~\ref{tetra},
is the simplest of the Platonic solids.
It consists of~4 trivalent vertices, 6 links and 4 triangular faces.

\begin{figure}[!ht]
\begin{center}
\includegraphics[angle=90,width=.3\textwidth]{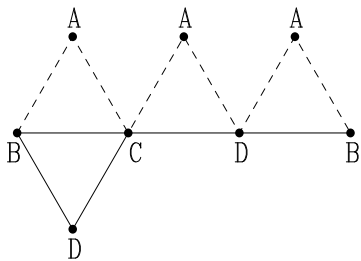}
\caption{
Planar representation of the tetrahedron.
Links drawn as dashed lines form the chosen spanning tree,
whereas those drawn as full lines carry non-trivial phase factors.}
\label{mtetra}
\end{center}
\end{figure}

Throughout the following it will be advantageous to unwrap
the polyhedra around an axis of high sym\-me\-try~\cite{sm}.
For the tetrahedron it is convenient to choose the threefold symmetry axis
going through A and the center of the opposite BCD face.
The planar representation thus obtained is shown in Figure~\ref{mtetra}.
Each face appears exactly once, whereas vertices and links may have
several occurrences, to be identified by the inverse procedure of wrapping
the planar representation onto the sphere.

In order to construct the gauge field
along the lines of the argument given in Section~\ref{general},
we have chosen a convenient, highly symmetric spanning tree,
shown in Figure~\ref{mtetra} as dashed lines.
The phase factors of these links are set to their trivial values $U_{ij}=1$.
The non-trivial phase factors are then obtained as follows.
The condition~(\ref{flux}) for
the faces ABC, ACD and ADB respectively yields
$U_{\rm BC}=\omega$, $U_{\rm CD}=\omega$ and $U_{\rm DB}=\omega$.
As far as the remaining face BCD is concerned,
using~(\ref{uustar}), one arrives at the relation $U_{\rm CB}U_{\rm
BD}U_{\rm DC}=\omega^{-3}$.
The condition~(\ref{flux}) for the face BCD is automatically fulfilled,
as a consequence of~(\ref{omegaf}) which reads $\omega^4=1$ since $F=4$,
so that $\omega^{-3}=\omega$.
This explicit example illustrates the general properties
that only $F-1$ of the $F$ conditions~(\ref{flux}) are independent.
To sum up, the trivial and non-trivial phase factors of the gauge field
respectively read
\beqa
&&U_{\rm AB}=U_{\rm AC}=U_{\rm AD}=1,\quad
U_{\rm BC}=U_{\rm CD}=U_{\rm DB}=\omega.
\eeqa

\begin{figure}[!ht]
\begin{center}
\includegraphics[angle=90,width=.4\textwidth]{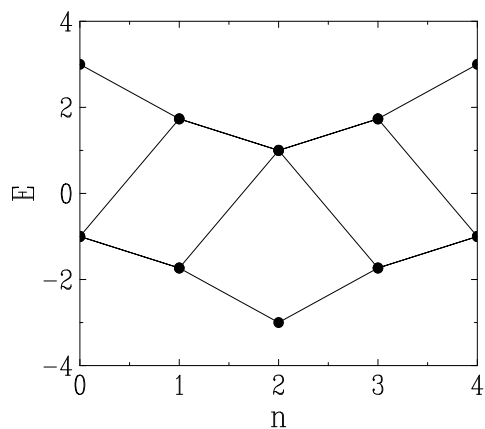}
\caption{
Energy levels $E$ of the tetrahedron as a function of the magnetic charge
$n$ over one period.
Line segments show how individual levels `jump'
as $n$ is increased by one unit.}
\label{stetra}
\end{center}
\end{figure}

The Hamiltonian matrix thus obtained,
\beq
\H=\pmatrix{0&1&1&1\cr 1&0&\omega&\omega^\star\cr
1&\omega^\star&0&\omega\cr 1&\omega&\omega^\star&0},
\eeq
depends on the magnetic charge~$n$ through the complex number $\omega$
introduced in~(\ref{omegadef}).
The energy eigenvalues of the above $4\times4$ matrix
are listed in Table~\ref{tabtetra} and shown in Figure~\ref{stetra}
for each value of the magnetic charge over one period.

\begin{table}[!ht]
\begin{center}
\begin{tabular}{|c|c|c|}
\hline
$n$&$E$&$m$\\
\hline
0&$\matrix{3\cr-1}$&$\matrix{1\cr3}$\\
\hline
1,3&$\matrix{\sqrt{3}^\es\cr-\sqrt{3}}$&$\matrix{2\cr2}$\\
\hline
2&$\matrix{1\cr-3}$&$\matrix{3\cr1}$\\
\hline
\end{tabular}
\end{center}
\caption{Energy levels $E$ of the tetrahedron and their multiplicities $m$,
for each value of the magnetic charge~$n$ over one period.}
\label{tabtetra}
\end{table}

The other Platonic solids will now successively
be dealt with along the same lines.

\subsection{The cube}
\label{subcube}

\begin{figure}[!ht]
\begin{center}
\includegraphics[angle=90,width=.3\textwidth]{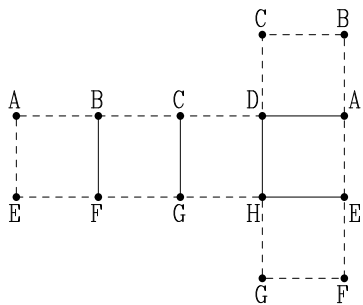}
\caption{
Planar representation of the cube.
Same convention as in Figure~\ref{mtetra}.}
\label{mcube}
\end{center}
\end{figure}

Figure~\ref{mcube} shows the planar representation
obtained by unwrapping the cube around a fourfold axis
going through the centers of the opposite faces ABCD and EFGH,
together with the spanning tree chosen to fix the gauge.
One has $\omega^6=1$.
The non-trivial phase factors of the gauge field read
\beqa
&&U_{\rm FB}=U_{\rm DA}=U_{\rm EH}=\omega,\quad
U_{\rm GC}=\omega^2,\quad U_{\rm HD}=\omega^3.
\eeqa

\begin{figure}[!ht]
\begin{center}
\includegraphics[angle=90,width=.4\textwidth]{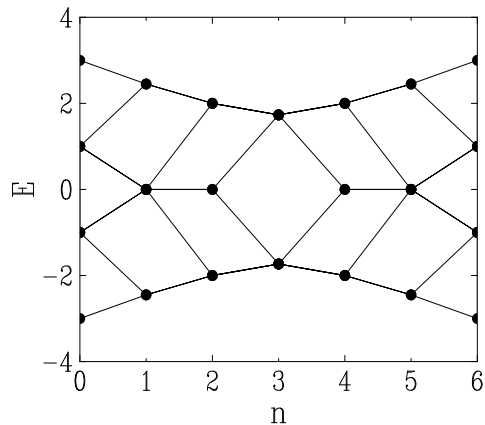}
\caption{
Energy levels $E$ of the cube as a function of the magnetic charge
$n$ over one period.
Same convention as in Figure~\ref{stetra}.}
\label{scube}
\end{center}
\end{figure}

The corresponding $8\times8$ Hamiltonian matrix
is too complex to be diagonalizable by hand.
For the cube and the subsequent Platonic solids,
this task has been performed with the help of the software MACSYMA.
The energy eigenvalues thus obtained
and their multiplicities are listed in Table~\ref{tabcube},
and shown in Figure~\ref{scube},
for each value of the magnetic charge over one period.

\begin{table}[!ht]
\begin{center}
\begin{tabular}{|c|c|c|}
\hline
$n$&$E$&$m$\\
\hline
0&$\matrix{3\cr1\cr-1\cr-3}$&$\matrix{1\cr3\cr3\cr1}$\\
\hline
1,5&$\matrix{\sqrt{6}^\es\cr0\cr-\sqrt{6}}$&$\matrix{2\cr4\cr2}$\\
\hline
\end{tabular}
\raisebox{16pt}{
\begin{tabular}{|c|c|c|}
\hline
$n$&$E$&$m$\\
\hline
2,4&$\matrix{2\cr0\cr-2}$&$\matrix{3\cr2\cr3}$\\
\hline
3&$\matrix{\sqrt{3}^\es\cr-\sqrt{3}}$&$\matrix{4\cr4}$\\
\hline
\end{tabular}}
\end{center}
\caption{Energy levels $E$ of the cube and their multiplicities $m$,
for each value of the magnetic charge~$n$ over one period.}
\label{tabcube}
\end{table}

\subsection{The octahedron}
\label{subocta}

\begin{figure}[!ht]
\begin{center}
\includegraphics[angle=90,width=.35\textwidth]{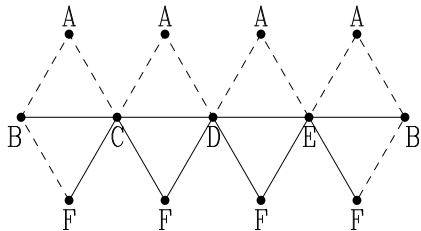}
\caption{
Planar representation of the octahedron.
Same convention as in Figure~\ref{mtetra}.}
\label{mocta}
\end{center}
\end{figure}

Figure~\ref{mocta} shows the planar representation
obtained by unwrapping the octahedron around a fourfold axis
going through the opposite vertices A and F,
together with the spanning tree chosen to fix the gauge.
One has $\omega^8=1$.
The non-trivial phase factors of the gauge field read
\beqa
&&U_{\rm BC}=U_{\rm CD}=U_{\rm DE}=U_{\rm EB}=\omega,\quad
U_{\rm EF}=U_{\rm FC}=\omega^2,\quad U_{\rm FD}=\omega^4.
\eeqa

\begin{figure}[!ht]
\begin{center}
\includegraphics[angle=90,width=.4\textwidth]{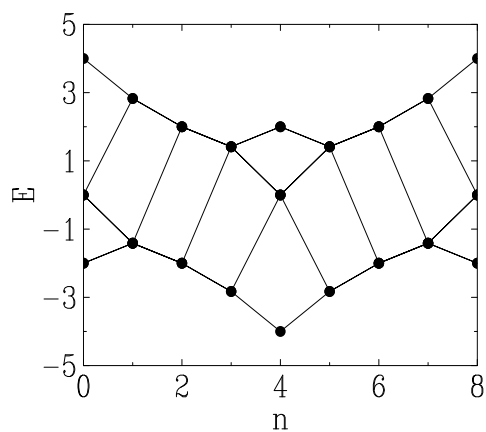}
\caption{
Energy levels $E$ of the octahedron as a function of the magnetic charge
$n$ over one period.
Same convention as in Figure~\ref{stetra}.}
\label{socta}
\end{center}
\end{figure}

The energy eigenvalues of the corresponding $6\times6$ Hamiltonian matrix
and their multiplicities are listed in Table~\ref{tabocta},
and shown in Figure~\ref{socta},
for each value of the magnetic charge over one period.

\begin{table}[!ht]
\begin{center}
\begin{tabular}{|c|c|c|}
\hline
$n$&$E$&$m$\\
\hline
0&$\matrix{4\cr0\cr-2}$&$\matrix{1\cr3\cr2}$\\
\hline
1,7&$\matrix{2\sqrt{2}^\es\cr-\sqrt{2}}$&$\matrix{2\cr4}$\\
\hline
2,6&$\matrix{2\cr-2}$&$\matrix{3\cr3}$\\
\hline
\end{tabular}
\raisebox{14.1pt}{
\begin{tabular}{|c|c|c|}
\hline
$n$&$E$&$m$\\
\hline
3,5&$\matrix{\sqrt{2}^\es\cr-2\sqrt{2}}$&$\matrix{4\cr2}$\\
\hline
4&$\matrix{2\cr0\cr-4}$&$\matrix{2\cr3\cr1}$\\
\hline
\end{tabular}}
\end{center}
\caption{Energy levels $E$ of the octahedron and their multiplicities $m$,
for each value of the magnetic charge~$n$ over one period.}
\label{tabocta}
\end{table}

\subsection{The dodecahedron}

\begin{figure}[!ht]
\begin{center}
\includegraphics[angle=90,width=.4\textwidth]{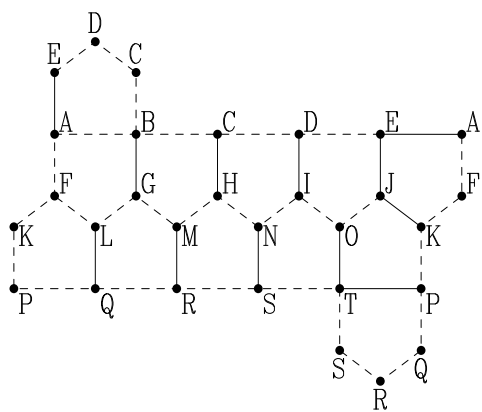}
\caption{
Planar representation of the dodecahedron.
Same convention as in Figure~\ref{mtetra}.}
\label{mdode}
\end{center}
\end{figure}

Figure~\ref{mdode} shows the planar representation
obtained by unwrapping the dodecahedron around a fivefold axis
going through the centers of the opposite faces ABCDE and PQRST,
together with the spanning tree chosen to fix the gauge.
One has $\omega^{12}=1$.
The non-trivial phase factors of the gauge field read
\beqa
&&U_{\rm EA}=U_{\rm GB}=U_{\rm QL}=U_{\rm PT}=\omega,\quad
U_{\rm HC}=U_{\rm RM}=\omega^2,\nonumber\\
&&U_{\rm ID}=U_{\rm SN}=\omega^3,\quad
U_{\rm JE}=U_{\rm TO}=\omega^4,\quad U_{\rm JK}=\omega^6.
\eeqa

\begin{figure}[!ht]
\begin{center}
\includegraphics[angle=90,width=.4\textwidth]{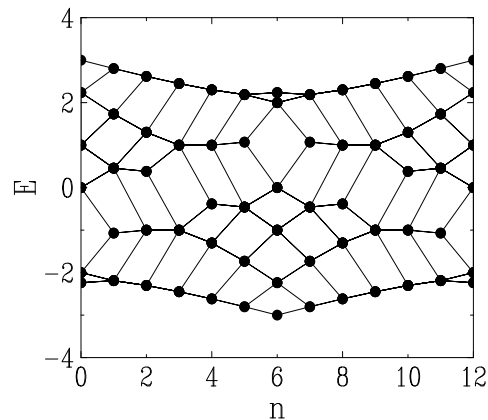}
\caption{
Energy levels $E$ of the dodecahedron as a function of the magnetic charge
$n$ over one period.
Same convention as in Figure~\ref{stetra}.}
\label{sdode}
\end{center}
\end{figure}

The energy eigenvalues of the corresponding $20\times20$ Hamiltonian matrix
and their multiplicities are listed in Table~\ref{tabdode},
and shown in Figure~\ref{sdode},
for each value of the magnetic charge over one period.

\begin{table}[!ht]
\begin{center}
\begin{tabular}{|c|c|c|}
\hline
$n$&$E$&$m$\\
\hline
0&$\matrix{3\cr\sqrt{5}\cr1\cr0\cr-2\cr-\sqrt{5}}$&$\matrix{1\cr3\cr5\cr4\cr4\cr3}$\\
\hline
1,11&$\matrix{\sqrt{3}(1+\sqrt{5}^\es)/2\cr\sqrt{3}\cr(\sqrt{7}-\sqrt{3})/2\cr\sqrt{3}(1-\sqrt{5})/2\cr-(\sqrt{7}+\sqrt{3})/2}$
&$\matrix{2\cr4\cr6\cr2\cr6}$\\
\hline
2,10&$\matrix{(3+\sqrt{5}^\es)/2\cr(\sqrt{13}-1)/2\cr(3-\sqrt{5})/2\cr-1\cr-(\sqrt{13}+1)/2}$
&$\matrix{3\cr5\cr3\cr4\cr5}$\\
\hline
3,9&$\matrix{\sqrt{6}^\es\cr1\cr-1\cr-\sqrt{6}}$&$\matrix{4\cr6\cr6\cr4}$\\
\hline
\end{tabular}
\raisebox{30.2pt}{
\begin{tabular}{|c|c|c|}
\hline
$n$&$E$&$m$\\
\hline
4,8&$\matrix{(\sqrt{13}^\es+1)/2\cr1\cr(\sqrt{5}-3)/2\cr(1-\sqrt{13})/2\cr-(3+\sqrt{5})/2}$
&$\matrix{5\cr4\cr3\cr5\cr3}$\\
\hline
5,7&$\matrix{(\sqrt{7}^\es+\sqrt{3})/2\cr\sqrt{3}(\sqrt{5}-1)/2\cr(\sqrt{3}-\sqrt{7})/2\cr-\sqrt{3}\cr-\sqrt{3}(1+\sqrt{5})/2}$
&$\matrix{6\cr2\cr6\cr4\cr2}$\\
\hline
6&$\matrix{\sqrt{5}^\es\cr2\cr0\cr-1\cr-\sqrt{5}\cr-3}$&$\matrix{3\cr4\cr4\cr5\cr3\cr1}$\\
\hline
\end{tabular}}
\end{center}
\caption{Energy levels $E$ of the dodecahedron and their multiplicities $m$,
for each value of the magnetic charge~$n$ over one period.}
\label{tabdode}
\end{table}

\subsection{The icosahedron}

\begin{figure}[!ht]
\begin{center}
\includegraphics[angle=90,width=.4\textwidth]{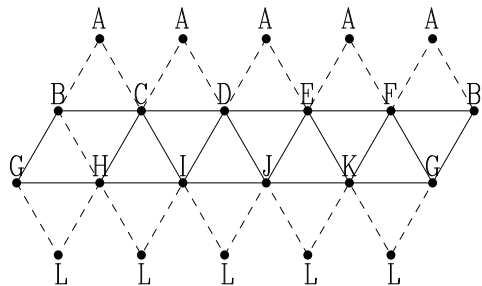}
\caption{
Planar representation of the icosahedron.
Same convention as in Figure~\ref{mtetra}.}
\label{mico}
\end{center}
\end{figure}

Figure~\ref{mico} shows the planar representation
obtained by unwrapping the icosahedron around a fivefold axis
going through the opposite vertices A and L,
together with the spanning tree chosen to fix the gauge.
One has $\omega^{20}=1$.
The non-trivial phase factors of the gauge field read
\beqa
&&U_{\rm BC}=U_{\rm CD}=U_{\rm DE}=U_{\rm EF}=U_{\rm FB}=\omega,\nonumber\\
&&U_{\rm GK}=U_{\rm KJ}=U_{\rm JI}=U_{\rm IH}=U_{\rm HG}=\omega,\nonumber\\
&&U_{\rm BG}=U_{\rm HC}=\omega^2,\quad
U_{\rm IC}=U_{\rm FG}=\omega^4,\quad
U_{\rm ID}=U_{\rm FK}=\omega^6,\nonumber\\
&&U_{\rm JD}=U_{\rm EK}=\omega^8,\quad
U_{\rm JE}=\omega^{10}.
\eeqa

\begin{figure}[!ht]
\begin{center}
\includegraphics[angle=90,width=.4\textwidth]{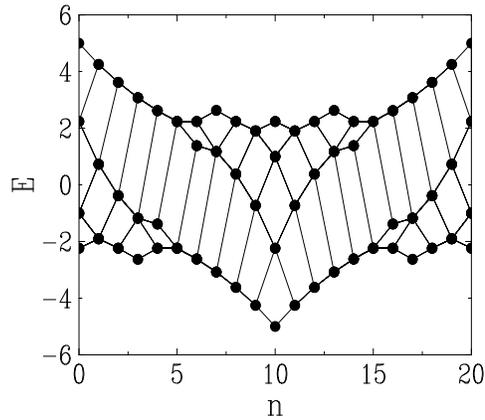}
\caption{
Energy levels $E$ of the icosahedron as a function of the magnetic charge
$n$ over one period.
Same convention as in Figure~\ref{stetra}.}
\label{sico}
\end{center}
\end{figure}

The energy eigenvalues of the corresponding $12\times12$ Hamiltonian matrix
and their multiplicities are listed in Table~\ref{tabico},
and shown in Figure~\ref{sico},
for each value of the magnetic charge over one period.

\begin{table}[!ht]
\begin{center}
\begin{tabular}{|c|c|c|}
\hline
$n$&$E$&$m$\\
\hline
0&$\matrix{5\cr\sqrt{5}\cr-1\cr-\sqrt{5}}$&$\matrix{1\cr3\cr5\cr3}$\\
\hline
1,19&$\matrix{\sqrt{10(5+\sqrt{5})}^\es/2\cr\sqrt{5-2\sqrt{5}}\cr-\sqrt{2(5+\sqrt{5})}/2}$&$\matrix{\raisebox{2pt}{2}\cr4\cr\raisebox{-2pt}{6}}$\\
\hline
2,18&$\matrix{(5+\sqrt{5}^\es)/2\cr(\sqrt{5}-3)/2\cr-\sqrt{5}}$&$\matrix{3\cr5\cr4}$\\
\hline
3,17&$\matrix{\sqrt{5+2\sqrt{5}}^\es\cr-\sqrt{2(5-\sqrt{5})}/2\cr-\sqrt{10(5-\sqrt{5})}/2}$&$\matrix{\raisebox{2pt}{4}\cr6\cr\raisebox{-2pt}{2}}$\\
\hline
4,16&$\matrix{(3+\sqrt{5}^\es)/2\cr(\sqrt{5}-5)/2\cr-\sqrt{5}}$&$\matrix{5\cr3\cr4}$\\
\hline
5,15&$\matrix{\sqrt{5}^\es\cr-\sqrt{5}}$&$\matrix{6\cr6}$\\
\hline
\end{tabular}
\raisebox{14.0pt}{
\begin{tabular}{|c|c|c|}
\hline
$n$&$E$&$m$\\
\hline
6,14&$\matrix{\sqrt{5}^\es\cr(5-\sqrt{5})/2\cr-(3+\sqrt{5})/2}$&$\matrix{4\cr3\cr5}$\\
\hline
7,13&$\matrix{\sqrt{10(5-\sqrt{5})}^\es/2\cr\sqrt{2(5-\sqrt{5})}/2\cr-\sqrt{5+2\sqrt{5}}}$&$\matrix{\raisebox{2pt}{2}\cr6\cr\raisebox{-2pt}{4}}$\\
\hline
8,12&$\matrix{\sqrt{5}^\es\cr(3-\sqrt{5})/2\cr-(5+\sqrt{5})/2}$&$\matrix{4\cr5\cr3}$\\
\hline
9,11&$\matrix{\sqrt{2(5+\sqrt{5})}^\es/2\cr-\sqrt{5-2\sqrt{5}}\cr-\sqrt{10(5+\sqrt{5})}/2}$&$\matrix{\raisebox{2pt}{6}\cr4\cr\raisebox{-2pt}{2}}$\\
\hline
10&$\matrix{\sqrt{5}^\es\cr1\cr-\sqrt{5}\cr-5}$&$\matrix{3\cr5\cr3\cr1}$\\
\hline
\end{tabular}}
\end{center}
\caption{Energy levels $E$ of the icosahedron and their multiplicities $m$,
for each value of the magnetic charge~$n$ over one period.}
\label{tabico}
\end{table}

\subsection{Properties of the spectra}
\label{props}

The energy spectra of the five Platonic solids have been given
in Tables~\ref{tabtetra} to~\ref{tabico}
and illustrated in Figures~\ref{stetra}, \ref{scube}, \ref{socta},
\ref{sdode} and~\ref{sico},
as a function of the magnetic charge~$n$ over one period.
These energy spectra exhibit several remarkable properties.
First of all,
the energy levels are {\it even} functions of the magnetic charge $n$,
i.e., they are invariant under a change of the sign of the magnetic field.
As announced above, they are also {\it periodic} in $n$, with period~$F$.

Another symmetry of the energy spectra can be revealed as follows.
Suppose all the (trivial and non-trivial)
phase factors $U_{ij}$ are simultaneously changed into their opposites.
The whole Hamiltonian matrix $\H$,
and therefore all the energy levels,
are also changed into their opposites.
On the other hand, as all the faces have $q$ sides,
the product in the left-hand side of~(\ref{flux}) is multiplied by
$(-1)^q$.
The magnetic flux through each face has therefore been increased by
$\delta\phi=q\Phi_0/2$,
i.e., $n$ has been increased by $qF/2$.
Therefore, if $q$ is even (this is the case for the cube),
the energy spectrum is its own opposite,
i.e., it is symmetric with respect to the origin of energies,
for every value of $n$.
If $q$ is odd (this is the case for the four other Platonic solids),
the spectrum at $n+F/2$ is the opposite of that at~$n$,
i.e., the spectrum is a {\it semi-periodic} function of~$n$.

All the energy levels have been found to be
algebraic numbers of degree at most four,
with relatively simple expressions involving square roots.
This property is somewhat surprising,
especially in view of the dimension of the Hamiltonian matrices,
which can be as high as $V=20$ for the dodecahedron.
The energy levels fulfill the sum rules~(\ref{sumrules}).
Checking this expected property is however a non-trivial
exercise in algebra for the case of the icosahedron.

The multiplicities of the energy levels are typically rather high,
and they vary in a complex and irregular fashion as a function
of both the magnetic charge and the rank of the considered energy level.
The spectra however exhibit the following property:
the largest energy eigenvalue has multiplicity $m=\abs{n}+1$
for a small enough absolute magnetic charge ($\abs{n}\le n_\max$,
where $n_\max=2,3,3,5,5$ for the five Platonic solids ordered as above).
This observed linear increase
is a perfect analogue of the continuum result~(\ref{mtamm}),
with the reasonable assumption that the continuum counterpart
of the largest energy eigenvalue corresponds to the orbital quantum number
$\ell=0$.

\subsection{Total energy}

An interesting illustration of the above energy spectra is provided
by the total energy at half filling, defined as
\beq
\E=\sum_{a=1}^{V/2} E_a,
\label{etotdef}
\eeq
where the $V$ energy levels are assumed to be in increasing order
($E_1\le E_2\le\dots\le E_V$)
and to be again repeated according to their multiplicities.
The first of the sum rules~(\ref{sumrules})
implies that the total energy $\E$ thus defined is insensitive to the
{\it sign} of the Hamiltonian~$\hat\H$,
i.e., it would be unchanged if the matrix
elements were replaced by their opposites $\H_{ij}=-U_{ij}$.
The second of the sum rules~(\ref{sumrules})
implies that the mean squared value of the individual energy levels reads
$\mean{E^2}=2L/V=p$, where $p$ is the coordination number of the vertices.
This observation is indicative that the total energy scales
as $\sqrt{p}\,V=\sqrt{2VL}$,
and therefore suggests to introduce the reduced total energy
\beq
\E_\rr=\frac{\E}{\sqrt{p}\,V}=\frac{\E}{\sqrt{2VL}}.
\label{erdef}
\eeq

This heuristic argument can be turned to a quantitative prediction
in the limit of an infinitely connected structure ($p\to\infty$),
such as e.g.~the hypercubic lattice in high spatial dimension $d$, where $p=2d$.
Although the full energy spectrum is the interval $-p\le E\le p$,
typical energy eigenvalues only grow as $E\sim\sqrt{p}$
in the limit under consideration,
and the {\it bulk} of the normalized density of states becomes Gaussian:
\beq
\rho_\infty(E)=\frac{\e^{-E^2/(2p)}}{\sqrt{2\pi p}}.
\eeq
The reduced total energy $\E_\rr$ therefore has
the following universal limiting value:
\beq
\E_\infty=\frac{1}{\sqrt{p}}\int_{-\infty}^0E\rho_\infty(E)\,\d E,
\eeq
i.e.,
\beq
\E_\infty=-\frac{1}{\sqrt{2\pi}}=-0.398942.
\label{elimit}
\eeq

\begin{figure}[!ht]
\begin{center}
\includegraphics[angle=90,width=.5\textwidth]{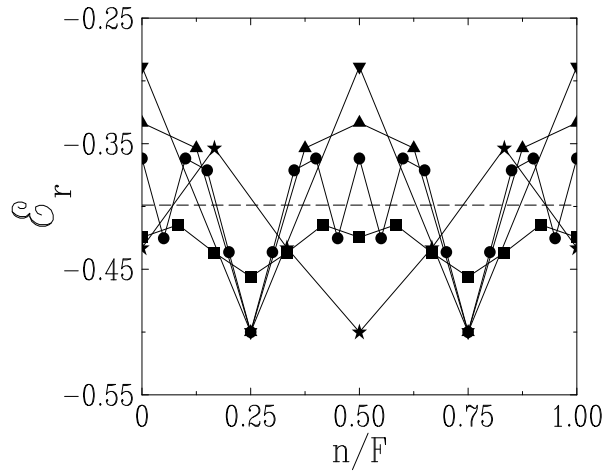}
\caption{
Plot of the reduced total energy $\E_\rr$
against the reduced flux per face $n/F$,
for the five Platonic solids:
tetrahedron (down triangles), cube (stars), octahedron (up triangles),
dodecahedron (squares) and icosahedron (circles).
The dashed line shows the limiting value~(\ref{elimit}).}
\label{etot}
\end{center}
\end{figure}

Figure~\ref{etot} shows a plot of the reduced total energy $\E_\rr$
as a function of the reduced flux per face,
$n/F=\phi/\Phi_0$, for the five Platonic solids.
This quantity is observed to vary within a rather
modest range around the limiting value~(\ref{elimit}), shown as a dashed line.

In the case of the cube, $\E_\rr$ is minimal for $n/F=1/2$,
in agreement with the prediction that the total energy is minimal
when the flux per face equals the filling factor~\cite{hase}.
In the four other cases, however, the last of the above symmetry properties
implies that $\E_\rr$ is periodic in $n$ with period $F/2$, and not $F$.
As a consequence, it takes the same value for $n/F=1/2$ and in the absence
of a magnetic field.
As it turns out,~$\E_\rr$ is observed to be minimal
for $n/F=1/4$ and $3/4$ ($F$ is a multiple of 4).

\section{The C$_{60}$ fullerene}
\label{sixty}

We now turn to the case of the C$_{60}$ fullerene.
This carbon molecule with the shape of a truncated icosahedron
has been discovered in 1985~\cite{Kroto}.
For simplicity we model it as a symmetric truncated icosahedron,
where all the links have equal lengths.
This symmetry is known to be slightly violated~\cite{lengths},
as for the free molecule
the length of the sides of the pentagons is 1.46~\AA,
whereas the length of the other links is 1.40~\AA.
Considering the symmetric polyhedron will however not affect
the salient qualitative features of our results.
It will indeed turn out that these features can be explained
by the quasiperiodic dependence of the energy spectrum on the magnetic charge.

The symmetric truncated icosahedron has $V=60$ equivalent vertices,
$L=90$ equivalent links, and $F=32$ faces, namely 12 pentagons and 20 hexagons,
respectively corresponding to the vertices and to the faces of the icosahedron.
Various results on its geometrical characteristics,
which are useful both in the present work and in~\cite{II},
are presented in a self-consistent fashion in the Appendix.
Figure~\ref{mfulle} shows the planar representation
obtained by unwrapping the fullerene around a fivefold axis
going through the opposite pentagonal faces
A$_1$A$_2$A$_3$A$_4$A$_5$ and L$_1$L$_2$L$_3$L$_4$L$_5$,
together with the spanning tree chosen to fix the gauge.
The labeling of the vertices is consistent with that used in Figure~\ref{mico}.

\begin{figure}[!ht]
\begin{center}
\includegraphics[angle=90,width=.7\textwidth]{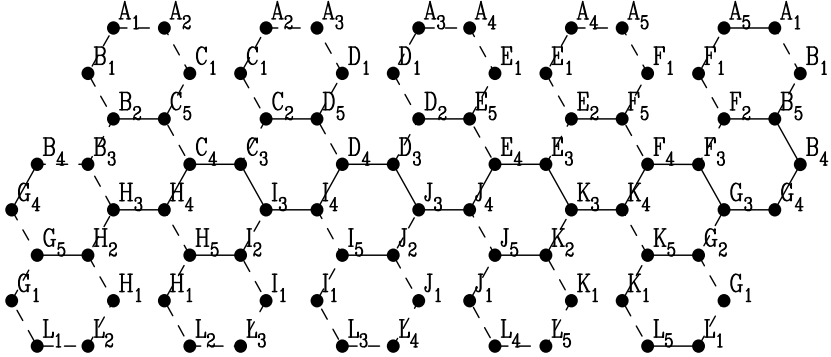}
\caption{
Planar representation of the fullerene.
Same convention as in Figure~\ref{mtetra}.
The 12 pentagonal faces A$_1\dots$A$_5$
to L$_1\dots$L$_5$ have not been drawn, for clarity.}
\label{mfulle}
\end{center}
\end{figure}

The magnetic fluxes $\phi_5$ and $\phi_6$
through a pentagonal and a hexagonal face
are proportional to the solid angles $\Omega_5$ and $\Omega_6$,
given by~(\ref{omega56}).
If $n$ denotes the magnetic charge,
the corresponding phase factors read
\beqa
&&\alpha=\exp(2\pi\ii\phi_5/\Phi_0)=\exp(\ii n\Omega_5/2),\nonumber\\
&&\beta=\exp(2\pi\ii\phi_6/\Phi_0)=\exp(\ii n\Omega_6/2).
\eeqa
Equation~(\ref{omegaid}) implies $\alpha^{12}\beta^{20}=1$.
The non-trivial phase factors of the gauge field are
\beqa
&&U_{\rm A_5A_1}=U_{\rm L_1L_5}=U_{\rm B_5B_4}=U_{\rm G_4G_3}
=U_{\rm C_4C_3}=U_{\rm D_4D_3}=\alpha,\nonumber\\
&&U_{\rm E_4E_3}=U_{\rm F_4F_3}
=U_{\rm H_4H_3}=U_{\rm I_4I_3}=U_{\rm J_4J_3}=U_{\rm K_4K_3}
=\alpha,\nonumber\\
&&U_{\rm B_2C_5}=U_{\rm C_2D_5}=U_{\rm D_2E_5}=U_{\rm E_2F_5}
=U_{\rm K_2J_5}=U_{\rm J_2I_5}=U_{\rm I_2H_5}=U_{\rm H_2G_5}
=\beta,\nonumber\\
&&U_{\rm F_2B_5}=U_{\rm G_2K_5}=\alpha\beta,\quad
U_{\rm B_4G_4}=\beta^2,\quad
U_{\rm H_4C_4}=\alpha\beta^2,\nonumber\\
&&U_{\rm I_3C_3}=\alpha^2\beta^4,\quad
U_{\rm I_4D_4}=\alpha^3\beta^6,\quad
U_{\rm J_3D_3}=\alpha^4\beta^8,\quad
U_{\rm J_4E_4}=\alpha^5\beta^{10},\nonumber\\
&&U_{\rm K_3E_3}=\alpha^6\beta^{12},\quad
U_{\rm K_4F_4}=\alpha^7\beta^{14},\quad
U_{\rm G_3F_3}=\alpha^9\beta^{16}.
\eeqa

\begin{figure}[!ht]
\begin{center}
\includegraphics[angle=90,width=.6\textwidth]{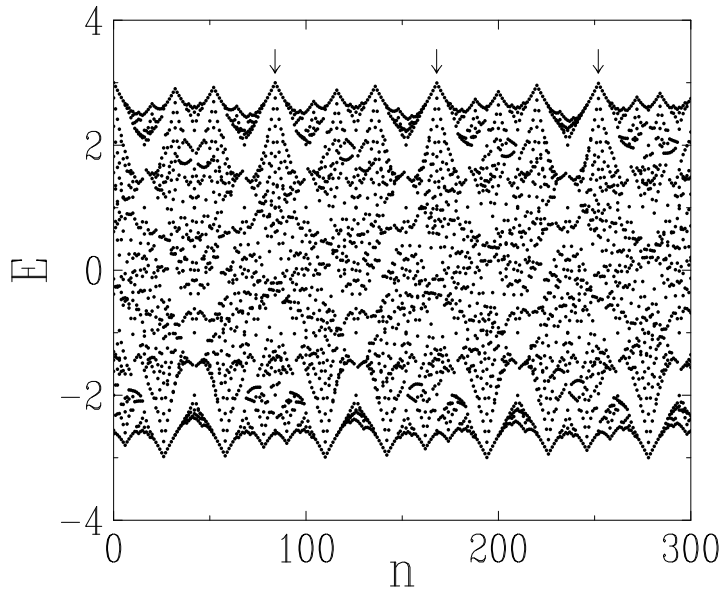}
\caption{
Energy spectrum of the fullerene as a function of the magnetic charge~$n$.
The arrows show the peaks at multiples of the quasi-period $Q=84$.}
\label{sfulle}
\end{center}
\end{figure}

Figure~\ref{sfulle} shows a plot of the
energy spectrum of the fullerene against the magnetic charge~$n$, up to $n=300$,
obtained by means of a numerical diagonalization of the Hamiltonian matrix.
In the absence of a magnetic monopole,
we recover the known tight-binding spectrum of the fullerene~\cite{manou},
with its 15 different levels having multiplicities ranging from 1 to 9.
For a non-zero magnetic charge~$n$,
the observed pattern of multiplicities only depends on the parity of $n$:

\begin{itemize}

\item
If $n$ is even, the spectrum consists of 16 distinct energy levels:
1 nondegenerate (with multiplicity 1),
6 with multiplicity 3, 4 with multiplicity 4 and 5 with multiplicity~5.

\item
If $n$ is odd, the spectrum consists of 14 distinct energy levels:
4 with multiplicity~2, 4 with multiplicity 4 and 6 with multiplicity 6.
All the multiplicities are even in this case.

Finally, the multiplicity of the largest energy eigenvalue
grows linearly with the absolute magnetic charge,
according to the rule $m=\abs{n}+1$,
already observed in the case of the Platonic solids,
up to $n_\max=5$ in the present case.

\end{itemize}

The Hamiltonian $\hat\H$, whose matrix elements involve
$\alpha$ and $\beta$, is a $4\pi$-periodic function
of $n\Omega_5$ and $n\Omega_6$, viewed as two independent variables.
In other words, it is a quasiperiodic function of the magnetic charge $n$,
because the solid angle ratio $R=\Omega_5/\Omega_6$,
given by~(\ref{omegaR}), is an irrational number.
The quasi-period $Q=84$, shown as arrows in Figure~\ref{sfulle},
clearly emerges from the pattern,
especially as sharp peaks where the highest energy level
is very close to its value $E=p=3$ in the absence of a magnetic charge.
The quasi-period $Q$ can be identified as the smallest integer
such that the reduced magnetic fluxes
through each type of face are very close to integers:
\beqa
&&\frac{\phi_5}{\Phi_0}=\frac{Q\Omega_5}{4\pi}=1.972412\approx2,
\nonumber\\
&&\frac{\phi_6}{\Phi_0}=\frac{Q\Omega_6}{4\pi}=3.016552\approx3.
\eeqa
The ratio $2/3$ is the first of the sequence of best rational approximants
to the solid angle ratio $R$ given in~(\ref{omegaR}).
The next approximant, $17/26$, corresponds to the quasi-period $Q=724$.

\begin{figure}[!ht]
\begin{center}
\includegraphics[angle=90,width=.4\textwidth]{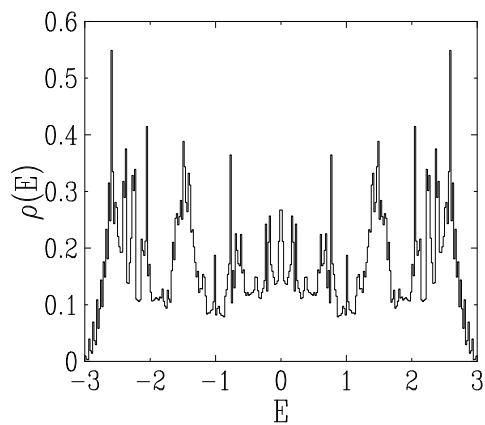}
\caption{
Histogram plot of the density of states $\rho(E)$ of the fullerene.
Spectra up to a magnetic charge $n=10^5$ are accumulated.
The bin size is $\delta E=0.02$.}
\label{fulledos}
\end{center}
\end{figure}

Figure~\ref{fulledos} shows a histogram plot of the density of states
obtained by accumulating the spectra up to $n=10^5$.
This procedure amounts to performing a uniform averaging over
the independent angles $n\Omega_5$ and $n\Omega_6$.
First of all, the density of states is observed to be an {\it even} function
of the energy $E$.
This property can be easily explained along the lines of Section~\ref{props}.
The Hamiltonian $\hat\H$ is indeed a semi-periodic function of $n\Omega_5$.

The very irregular behavior of the density of states,
with its many narrow peaks
whose height keeps growing as the bin size $\delta E$ is decreased,
is reminiscent of the singular spectra of more conventional
quasiperiodic structures, especially in low dimension,
such as the Fibonacci chain (see~\cite{fibospectrum} for a recent review).
It is, however, worth noticing that the density of states
also has a smooth and rather uniform background
all over the spectrum, i.e., for $-3\le E\le3$.

An intriguing question is whether the study of the fullerene spectrum
can tell us something about the spectrum of graphene~\cite{Guinea}.
In the absence of magnetic field, the density of states of graphene
vanishes linearly for energies close to half filling~\cite{graphene_review},
as $\rho(E)\propto\abs{E}$,
whereas in the presence of a magnetic field $B$ there are Landau levels
at $E_m={\rm sign}(m)\sqrt{2e\hbar v^2B|m|}$ ($m=0, \pm1, \pm2, \dots$),
where $v$ is the band velocity~\cite{LLG}.
In particular, there is a Landau level at zero energy, as $E_0=0$.
It is, however, too speculative to interpret the peak at $E=0$
visible in Figure~\ref{fulledos} as a zero-energy Landau level.

\begin{figure}[!ht]
\begin{center}
\includegraphics[angle=90,width=.44\textwidth]{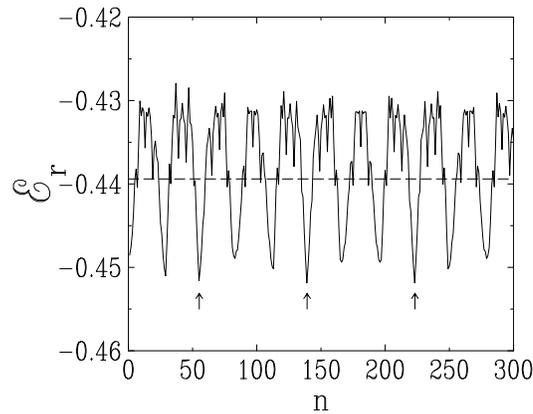}
\caption{
Plot of the reduced total energy $\E_\rr$ of the fullerene
against the magnetic charge $n$.
The arrows show that the minima are separated by multiples
of the quasi-period $Q=84$.
The dashed line shows the mean $\mean{\E_\rr}=-0.4394$.}
\label{fulletot}
\end{center}
\end{figure}

Figure~\ref{fulletot} shows a plot of the reduced total energy at half filling,
$\E_\rr=\E/(60\sqrt{3})$ against the magnetic charge~$n$, up to $n=300$.
The data exhibit oscillations at the above quasi-period $Q=84$,
with a strong third harmonic, especially near the minima.
The mean reduced total energy, $\mean{\E_\rr}=-0.4394$,
is only 10.2 percent larger (in absolute value) than the limiting value
$E_\infty$ of~(\ref{elimit}),
although the density of states shown in Figure~\ref{fulledos}
is very far from being a Gaussian.

\section{The diamond and the prism}
\label{last}

We end up this investigation by considering the following
two families of polyhedra.

\begin{itemize}

\item
The {\it diamond} is obtained by gluing together
two pyramids whose basis is a regular polygon with $N$ sides.
It has $N+2$ vertices, the two poles (with coordination number $N$)
and $N$ vertices along the polygonal equator (with coordination number 4),
$3N$ links and $2N$ equivalent triangular faces.

\item
The {\it prism}
is a rectangular prism whose basis is a regular polygon with $N$ sides.
It has $2N$ equivalent vertices with coordination number 3,
$3N$ links,
and $N+2$ faces, the two polygonal bases and $N$ lateral rectangular faces.

\end{itemize}

\begin{figure}[!ht]
\begin{center}
\includegraphics[angle=90,width=.4\textwidth]{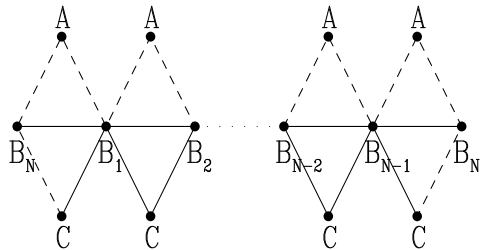}
\caption{
Planar representation of the diamond.
Same convention as in Figure~\ref{mtetra}.}
\label{mdiamond}
\end{center}
\end{figure}

\begin{figure}[!ht]
\begin{center}
\includegraphics[angle=90,width=.4\textwidth]{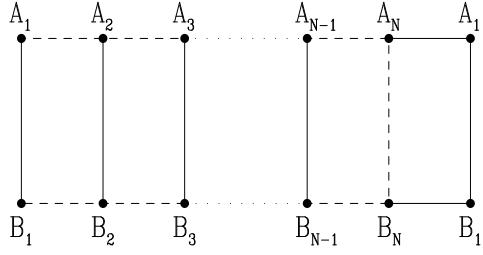}
\caption{
Planar representation of the prism.
Same convention as in Figure~\ref{mtetra}.
The upper and lower polygonal bases A$_1\dots$A$_N$
and B$_1\dots$B$_N$ have not been drawn, for clarity.}
\label{mprism}
\end{center}
\end{figure}

There is an infinite family of each kind of polyhedra,
labeled by the integer $N\ge3$.
The $N$-prism and the $N$-diamond are {\it dual} to each other
(see e.g.~\cite{wilson}).
Two of the Platonic solids, namely the octahedron and the cube,
are recovered for $N=4$.
Figures~\ref{mdiamond} and~\ref{mprism} respectively show
the planar representations
obtained by unwrapping the diamond and the prism around their $N$-fold axis,
together with the spanning trees chosen to fix the gauge.
The tight-binding spectra will be worked out explicitly
for each family of graphs, for arbitrary $N$ and an arbitrary magnetic charge.

\subsection{The diamond}

The $2N$ triangular faces of the diamond are equivalent.
As a consequence, if the magnetic charge is $n$,
the magnetic flux per face is $n\Phi_0/(2N)$,
so that the right-hand side of~(\ref{flux}) reads
\beq
\omega=\exp(\ii\pi n/N)
\eeq
and obeys
\beq
\omega^N=(-1)^n,\quad\omega^{2N}=1.
\eeq
The latter property ensures that the problem is periodic in the integer $n$,
with period~$2N$.

With the notations of Figure~\ref{mdiamond},
the non-trivial phase factors of the gauge field read
\beq
U_{\rm B_{\mathit k}B_{{\mathit k}+1}}=\omega,\quad
U_{\rm CB_{\mathit k}}=\omega^{2k}
\eeq
for $k=1,\dots,N$, with periodic boundary conditions
($N+1\equiv1$).
The eigenvalue equation~(\ref{eigen}) therefore reads
\beqa
&&E\psi_k=\omega\psi_{k-1}+\omega^{-1}\psi_{k+1}
+\psi_{\rm A}+\omega^{2k}\psi_{\rm C},\nonumber\\
&&E\psi_{\rm A}=\sum_{k=1}^N\psi_k,\quad
E\psi_{\rm C}=\sum_{k=1}^N\omega^{-2k}\psi_k,
\eeqa
where the $\psi_k=\psi_{\rm B_{\mathit k}}$ obey periodic boundary conditions.
Setting $a_k=\omega^k\psi_k$, the above equations simplify to
\beqa
&&Ea_k=a_{k-1}+a_{k+1}
+\omega^{-k}\psi_{\rm A}+\omega^k\psi_{\rm C},\nonumber\\
&&E\psi_{\rm A}=\sum_{k=1}^N\omega^ka_k,\quad
E\psi_{\rm C}=\sum_{k=1}^N\omega^{-k}a_k,
\label{ed}
\eeqa
where the $a_k$ obey the boundary conditions $a_0=\omega^Na_N$,
$a_{N+1}=\omega^Na_1$.
The eigenvalues and eigenfunctions solving~(\ref{ed})
can be derived explicitly as follows.

Consider first a generic value of the magnetic charge ($n\ne0$ and $n\ne N$),
so that $\omega$ is not real,
and look for an extended plane-wave solution
of the form $a_k=\e^{\ii kq}$, so that $E=2\cos q$.
The boundary conditions yield $q=(2m-n)\pi/N$, with $m=1,\dots,N$.
One has $\psi_{\rm A}=\psi_{\rm C}=0$, which imposes $m\ne n$ and $m\ne N$.
We thus obtain a band of $N-2$ eigenvalues,
\beq
E=2\cos\frac{(2m-n)\pi}{N},
\label{band}
\eeq
with a hole at
\beq
E_0=2\cos\frac{n\pi}{N},
\label{hole}
\eeq
corresponding to $m=n$ being forbidden.
Four other solutions correspond to impurity states:
either $a_k=\omega^k$, so that $\psi_{\rm A}=0$ but $\psi_{\rm C}\ne0$,
or $a_k=\omega^{-k}$, so that $\psi_{\rm C}=0$ but $\psi_{\rm A}\ne0$.
Both cases yield two minibands of twice degenerate impurity levels:
\beq
E=\cos\frac{n\pi}{N}\pm\sqrt{N+\cos^2\frac{n\pi}{N}}.
\label{mini}
\eeq

Consider now the values $n=0$ and $n=N$ of the magnetic charge,
respectively corresponding to an integer and a half-integer flux quantum
per face, i.e., to $\omega=1$ and $\omega=-1$.
There are now $N-1$ eigenvalues in the band~(\ref{band}), which has no
hole.
Three other solutions correspond to impurity states.
First, $a_k=0$ and $\psi_{\rm A}+\psi_{\rm C}=0$
yields one impurity state right at the band center, i.e., at energy
\beq
E=0.
\label{isoin}
\eeq
Second, $a_k=\omega^k$ and $\psi_{\rm A}=\psi_{\rm C}$
yields two nondegenerate impurity levels:
\beq
E=\omega\pm\sqrt{2N+1},
\label{isoout}
\eeq
lying further away from the band than the minibands of~(\ref{mini}).
The energy spectrum thus obtained can be checked to obey
the sum rules~(\ref{sumrules}) with $L=3N$,
and to give back the spectrum of the octahedron
(see Section~\ref{subocta}) in the case $N=4$.

\begin{figure}[!ht]
\begin{center}
\includegraphics[angle=90,width=.5\textwidth]{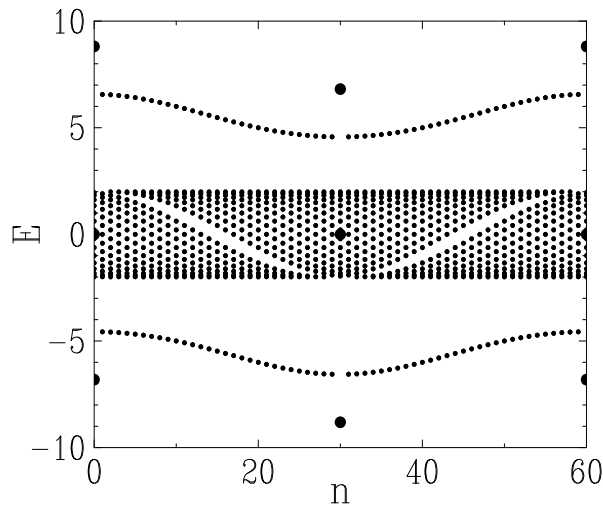}
\caption{
Energy levels $E$ of the diamond for $N=30$ as a function
of the magnetic charge~$n$ over one period.
Large dots show the isolated impurity levels (\ref{isoin}) and (\ref{isoout}).}
\label{sdiamond}
\end{center}
\end{figure}

Figure~\ref{sdiamond} shows a plot of the energy levels
of the diamond for $N=30$ as a function of the magnetic charge~$n$
over one period.
The main features of the spectrum are clearly visible:
the band~(\ref{band}) of extended states with its hole~(\ref{hole}),
the minibands~(\ref{mini}) of impurity states,
and the isolated impurity levels in the two commensurate cases,
shown as larger dots,
namely~(\ref{isoin}) within the band and~(\ref{isoout}) away from the
band.

\subsection{The prism}

The prism has two different types of faces,
two polygonal bases and $N$ rectangular faces.
The solid angles of each type of face viewed from the center of the solid
could be evaluated explicitly,
using results from the Appendix, as a function of $N$ and of the aspect ratio
$h/a$, where $h$ is the height of the prism
and $a$ the side of its polygonal bases.
This parametrization will not be needed in the following.

Let $n$ be the magnetic charge,
$\phi_{\rm B}$ the flux through either of the polygonal bases,
and $\phi$ the flux through a rectangular face.
One has then $2\phi_{\rm B}+N\phi=n\Phi_0$.
The corresponding phase factors,
\beq
\alpha=\exp(2\pi\ii\phi_{\rm B}/\Phi_0),\quad\beta=\exp(2\pi\ii\phi/\Phi_0),
\eeq
obey
\beq
\alpha^2\beta^N=1.
\eeq
In the following we take $\beta$ as our basic variable,
at fixed magnetic charge~$n$,
forgetting about its dependence on the geometry of the prism.
We introduce for further convenience the variable
\beq
u=\frac{2\pi\phi}{\Phi_0},
\eeq
such that $\beta=\exp(\ii u)$.

With the notations of Figure~\ref{mprism},
the non-trivial phase factors of the gauge field read
\beq
U_{\rm A_{\mathit N}A_1}=U_{\rm B_1B_{\mathit N}}=\alpha,\quad
U_{\rm B_{\mathit k}A_{{\mathit k}}}=\beta^k.
\eeq
The eigenvalue equation~(\ref{eigen}) therefore reads
\beqa
&&Ea_k=a_{k-1}+a_{k+1}+\beta^kb_k,\quad
Eb_k=b_{k-1}+b_{k+1}+\beta^{-k}a_k,
\eeqa
where the $a_k=\psi_{\rm A_{\mathit k}}$ and $b_k=\psi_{\rm B_{\mathit k}}$
obey the boundary conditions
$a_0=\alpha a_N$, $a_1=\alpha a_{N+1}$, $b_N=\alpha b_0$, $b_{N+1}=\alpha b_1$.
Setting $c_k=\beta^kb_k$, the above equations simplify to
\beqa
&&Ea_k=a_{k-1}+a_{k+1}+c_k,\quad
Ec_k=\beta c_{k-1}+\beta^{-1}c_{k+1}+a_k,
\label{ep}
\eeqa
whereas the $c_k$ obey the same boundary conditions as the $a_k$, i.e.,
$c_0=\alpha c_N$, $c_1=\alpha c_{N+1}$.
The eigenvalues and eigenfunctions solving~(\ref{ep})
can be derived explicitly as follows.
Looking for an extended plane-wave solution of the form
\beq
\pmatrix{a_k\cr c_k}=\pmatrix{\lambda\cr\mu}e^{\ii kq},
\eeq
Equation~(\ref{ep}) implies that the energy $E$ fulfills the condition
\beq
\left\vert\matrix{2\cos q-E&1\cr1&2\cos(q-u)-E}\right\vert=0.
\eeq
Introducing the shifted momentum
\beq
Q=q-\frac{u}{2}=q-\frac{\pi\phi}{\Phi_0},
\eeq
the above equation yields the following two-band dispersion relation
\beq
E_\pm(Q)=2\cos Q\cos\frac{u}{2}\pm\sqrt{1+4\sin^2Q\sin^2\frac{u}{2}}.
\eeq
The boundary conditions lead to the following
$N$ quantized values of the momentum:
\beq
Q=\frac{(2m+n)\pi}{N}\quad(m=1,\dots,N).
\label{qcon}
\eeq

\begin{figure}[!ht]
\begin{center}
\includegraphics[angle=90,width=.5\textwidth]{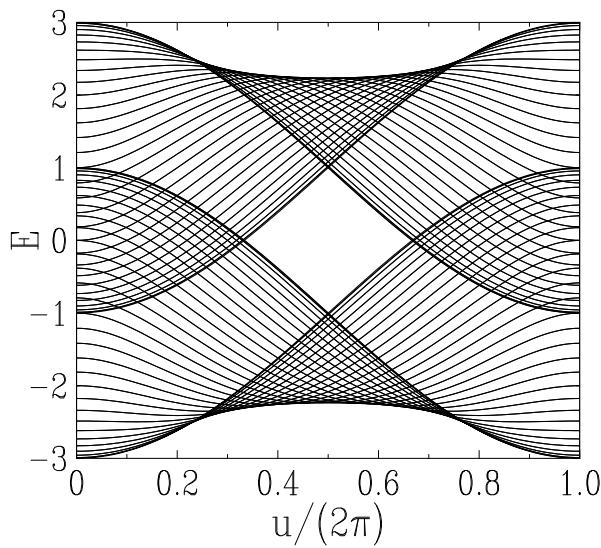}
\caption{
Energy levels $E$ of the prism for $N=30$
against $u/(2\pi)=\phi/\Phi_0$ over one period,
for an even magnetic charge~$n$.}
\label{sprism}
\end{center}
\end{figure}

Each energy level is an even and periodic function,
whose period is, as expected,~$2\pi$ in $u$, i.e., $\Phi_0$ in $\phi$.
Furthermore, the spectrum bears an extra discrete dependence
on the magnetic charge~$n$ through the quantization condition~(\ref{qcon}).
In fact, only the parity of $n$ matters,
and levels for even $n$ and odd $n$ alternate.
Figure~\ref{sprism} shows a plot of the energy spectrum of the prism
with $N=30$ against $u/(2\pi)$ over one period,
for an even magnetic charge~$n$.
The levels for odd $n$ would lie between these levels.
The energy spectrum thus obtained can be checked to obey
the sum rules~(\ref{sumrules}) with $L=3N$,
and to give back the spectrum of the cube
(see Section~\ref{subcube}) for $N=4$ and $u=n\pi/3$.

\begin{figure}[!ht]
\begin{center}
\includegraphics[angle=90,width=.45\textwidth]{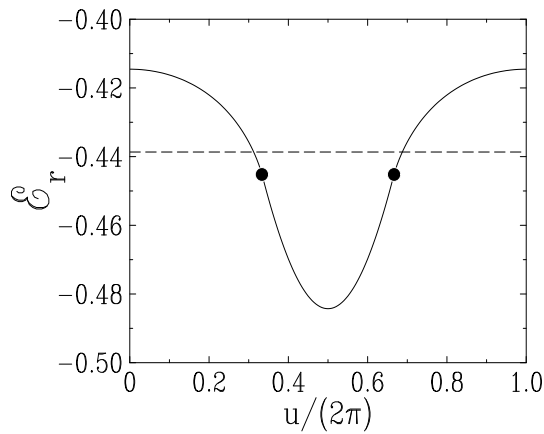}
\caption{
Plot of the limit $\E_\rr(u)$
of the reduced total energy of the prism,
against $u/(2\pi)=\phi/\Phi_0$ over one period.
Dots show the values of $u$ which separate the regions
where the bands overlap or not,
so that $\E_\rr(u)$ has different analytic expressions.
The dashed line shows the mean $\mean{\E_\rr}=-0.43866$.}
\label{prismtot}
\end{center}
\end{figure}

The reduced total energy at half filling, defined in~(\ref{erdef}),
has a well-defined limit $\E_\rr(u)$ as the number of sites
of the prism becomes infinitely large,
irrespective of the parity of the magnetic charge~$n$.
This limiting function has two different expressions,
involving incomplete elliptic integrals,
according to whether the bands $E_\pm(Q)$ overlap
($u<2\pi/3$ and $u>4\pi/3$) or not ($2\pi/3<u<4\pi/3$).
Figure~\ref{prismtot} shows a plot of $\E_\rr(u)$ against $u/(2\pi)$.
The mean reduced total energy, $\mean{\E_\rr}=-0.43866$,
is only 9.9 percent larger (in absolute value) than the limiting value
$\E_\infty$ of~(\ref{elimit}).

\section{Discussion}
\label{discussion}

We have presented an extensive study of the tight-binding spectra
on various polyhedral graphs drawn on the unit sphere,
as a function of the magnetic field produced by a quantized
magnetic charge sitting at the center of the sphere.
For a fixed polyhedron, there is only one discrete parameter left,
namely the integer magnetic charge~$n$, fixing the gauge sector of the model.

The spectra of the five Platonic solids, described in Section~\ref{plato},
exhibit a periodic dependence on the magnetic charge~$n$,
the period being the number of faces $F$ of the polyhedron.
The observed multiplicities of the energy levels are typically high.
The multiplicity $m=\abs{n}+1$ of the largest energy eigenvalue,
observed for the smallest values of the absolute magnetic charge,
is a discrete analogue of the known continuum spectrum
of multiplicities dating back to Tamm~\cite{tamm}.
All the energy levels have rather simple expressions,
involving at most two square roots,
although the dimension of the Hamiltonian matrices can be as large as
20 in the case of the dodecahedron.
It would be worthwhile to explore this rich pattern of multiplicities
within a more formal group-theoretical framework.
The case of the C$_{60}$ fullerene,
modeled as a symmetric truncated icosahedron,
is dealt with in Section~\ref{sixty}.
The main features of the spectrum are explained
by the quasiperiodic dependence of the energy spectrum
on the magnetic charge~$n$,
and mainly by the occurrence of a quasi-period $Q=84$,
related to the first rational approximant of the area ratio.
These features would still be present in the generic case
of a non-symmetric truncated icosahedron, with two different bond lengths.
Finally, the two families of polyhedra investigated in Section~\ref{last}
have given to us the opportunity of underlining yet other features
of the problem pertaining to less symmetric situations,
such as the relevance of one-dimensional band structures
in the very anisotropic regimes of the prism and the diamond at large~$N$.

The investigation of physical properties has been focused onto
the total energy $\E$ at half filling.
For all the polyhedra where all the vertices
have the same coordination number $p$
(i.e., all the examples considered in this work except for the diamond),
the typical value of the total energy is found to be rather close
(within a range of ten to twenty percent)
to its universal large-$p$ behavior corresponding to the asymptotically
Gaussian nature of the density of states.
The behavior of the total energy as a function of the magnetic charge~$n$
depends on the underlying polyhedron.
The total energy is minimal when the magnetic flux per face
equals $1/2$ flux quantum for the cube, and more generally for the
family of prisms, in agreement with the prediction that the minimum
occurs when the flux per face equals the filling factor~\cite{hase}.
In the other four polyhedra, to the contrary,
two energy minima are attained for the fractions $1/4$ and $3/4$,
as a consequence of an extra symmetry of the spectra.

\subsection*{Acknowledgments}

It is a pleasure for us to thank S.I.~Ben-Abraham, B.~Dou\c cot
and G.~Montambaux for very stimulating discussions.

\appendix
\section{The C$_{60}$ fullerene.~Geometrical characteristics}

For simplicity we model the C$_{60}$ fullerene
as a symmetric truncated icosahedron, where all the links have equal lengths.
This polyhedron consists of 12 pentagons and 20 hexagons,
respectively corresponding to the vertices and to the faces of the icosahedron.
The goal of this Appendix is to derive in a self-contained fashion
expressions for some of its geometrical characteristics,
to be used both in this work and in~\cite{II},
and especially the solid angles $\Omega_5$ and $\Omega_6$.

\begin{figure}[!ht]
\begin{center}
\includegraphics[angle=90,width=.2\textwidth]{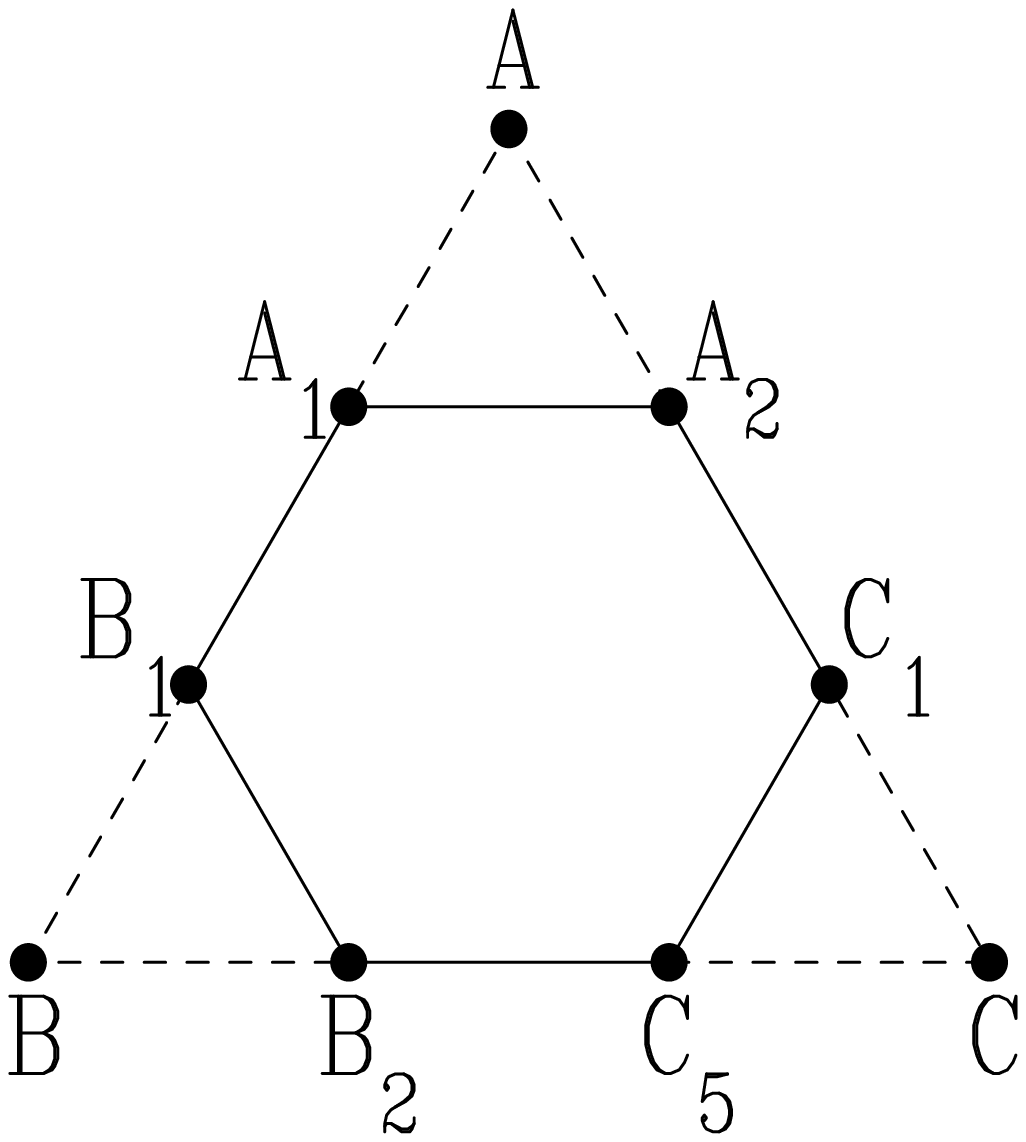}
\caption{
The triangular face ABC of the icosahedron
decorated by vertices of the fullerene.
Notations are consistent with Figures~\ref{mico} and~\ref{mfulle}.}
\label{tri}
\end{center}
\end{figure}

Figure~\ref{tri} shows an enlargement of the upper left part
of Figures~\ref{mico} and~\ref{mfulle}, with consistent notations.
We denote by $\A$ the vector from the origin to the vertex A, and so on.
Choosing for convenience a coordinate system such that A is at the North pole
and~B lies in the $xz$-plane, the coordinates of the vertices
A, B and C of the icosahedron~read
\beqa
&&{\hskip -10pt}\A=(0,0,1),\nonumber\\
&&{\hskip -10pt}\B=(\sin\theta_\ii,0,\cos\theta_\ii),\nonumber\\
&&{\hskip
-10pt}\C=(\sin\theta_\ii\cos(2\pi/5),\sin\theta_\ii\sin(2\pi/5),\cos\theta_\ii),
\eeqa
Let $\theta_\ii$ be the arc length of the links of the icosahedron.
One has $\cos\theta_\ii=\A\cdot\B=\B\cdot\C$, hence
\beq
\cos\theta_\ii=\frac{1}{\sqrt{5}}.
\label{thi}
\eeq

The vertices A$_1$, A$_2$, B$_1$ and C$_1$
of the fullerene can be parametrized as
\beqa
&&\A_1=\lambda\B+\mu\A,\quad\A_2=\lambda\C+\mu\A,\nonumber\\
&&\B_1=\lambda\A+\mu\B,\quad\C_1=\lambda\A+\mu\C.
\eeqa
The conditions $\A_1^2=1$ and $\A_1\cdot\A_2=\A_1\cdot\B_1$ yield
$\mu=2\lambda$ and
\beq
\lambda=\sqrt\frac{25-4\sqrt{5}}{109}.
\eeq
Let $\theta_\f$ be the arc length of the links of the fullerene.
One has
\beq
\cos\theta_\f=\A_1\cdot\A_2=(4+\sqrt{5})\lambda^2=\frac{80+9\sqrt{5}}{109}.
\label{cth}
\eeq

The above expressions can be used to evaluate the relevant solid angles.
Consider a spherical triangle defined by three points A, B, C on the unit
sphere.
It is well-known that the area (solid angle) of the triangle is
given by the spherical excess
$\Omega=\hat{\mathrm A}+\hat{\mathrm B}+\hat{\mathrm C}-\pi$,
where $\hat{\mathrm A}$ is the angle of the triangle at its vertex A, and so on.
This result is however not very useful in the present situation,
where the vertices are known through their coordinates.
A more convenient expression reads~\cite{kk}
\beq
\tan^2\frac{\Omega}{4}
=\tan\frac{s}{2}\tan\frac{s-a}{2}\tan\frac{s-b}{2}\tan\frac{s-c}{2},
\label{tankk}
\eeq
where $a$, $b$, $c$ are the arc lengths of the sides,
such that $\cos a=\B\cdot\C$, and so on,
and $s=(a+b+c)/2$ is half the perimeter.
The formula~(\ref{tankk}) is the spherical analogue
of a well-known expression for the area of a planar triangle,
\beq
{\cal A}^2=s(s-a)(s-b)(s-c),
\label{heron}
\eeq
dating back to Antiquity and known as Heron's formula.
Equation~(\ref{tankk}) can therefore be referred to as the spherical Heron formula.
It can be recast, using trigonometric identities,
into the following alternative form~\cite{edm}:
\beq
\cos\frac{\Omega}{2}
=\frac{\cos^2\frad{a}{2}+\cos^2\frad{b}{2}+\cos^2\frad{c}{2}-1}
{2\cos\frad{a}{2}\cos\frad{b}{2}\cos\frad{c}{2}}
=\frac{\cos a+\cos b+\cos c+1}
{4\cos\frad{a}{2}\cos\frad{b}{2}\cos\frad{c}{2}}.
\eeq
Some further algebra led us to the following form,
that is especially convenient in the case where the vertices
are known through their Cartesian coordinates:
\beq
1-\cos\Omega=\frac{(\A,\B,\C)^2}{(1+\A\cdot\B)(1+\B\cdot\C)(1+\C\cdot\A)},
\label{coskk}
\eeq
where
\beq
(\A,\B,\C)=(\A\times\B)\cdot\C=\A\cdot(\B\times\C)
\eeq
is the scalar triple product of the three unit vectors.

For the isosceles triangle AA$_1$A$_2$, the formula~(\ref{coskk}) yields
after some algebra
\beq
\cos\Omega_{\rm T}
=\frac{13-4\sqrt{5}+\sqrt{605+184\sqrt{5}}}{36}.
\eeq
The solid angles of a pentagonal and a hexagonal face obey
\beq
12\Omega_5+20\Omega_6=4\pi
\label{omegaid}
\eeq
and read
\beq
\Omega_5=5\Omega_{\rm T},\quad
\Omega_6=\frac{\pi}{5}-3\Omega_{\rm T}.
\eeq
Some further algebra using trigonometric identities yields
\beq
\cos\Omega_5=\frac{54887+720\sqrt{5}}{59049},\quad
\cos\Omega_6=\frac{511+945\sqrt{5}}{2916},
\label{omega56}
\eeq
hence the numerical values
\beq
\Omega_5=0.295072,\quad\Omega_6=0.451275.
\eeq

To close up, it is worth comparing the ratio
\beq
R=\frac{\Omega_5}{\Omega_6}=0.653863
\label{omegaR}
\eeq
to its planar analogue,
namely the area ratio of a pentagon and a hexagon with the same side.
In the plane, the area of a regular $n$-gon of side $a$ is
\beq
{\cal A}_n=\frac{na^2}{4\tan(\pi/n)}.
\eeq
The area ratio therefore reads
\beq
r=\frac{{\cal A}_5}{{\cal A}_6}=\frac{\sqrt{15(5+2\sqrt{5})}}{18}=0.662212.
\label{omegar}
\eeq
The comparison of both results~(\ref{omegaR}) and~(\ref{omegar})
shows that the effect of curvature is rather weak, as one has $R/r=0.987392$.


\section*{References}

\begin{thebibliography}{99}

\bibitem{Imry} Imry Y, 2002 {\it Introduction to Mesoscopic Physics} 2nd ed
(Oxford: Oxford University Press)

\bibitem{Giamarchi} Giamarchi T, 2004 {\it Quantum Physics in one dimension}
(Oxford: Oxford University Press)

\bibitem{Vignale} Giuliani G F and Vignale G, 2005 {\it Quantum Theory of the
Electron Liquid} (New York: Cambridge University Press)

\bibitem{II} Avishai Y and Luck J M, 2008 {\it Tight-binding electronic spectra on graphs with spherical topology II: the effect of spin-orbit interaction} Preprint arXiv:0802.0795

\bibitem{Kroto} Kroto H W, Heath J R, O'Brien S C, Curl R F and Smalley R E, 1985 {\it Nature} {\bf 318} 162

\bibitem{tamm} Tamm I, 1931 {\it Z. Phys.} {\bf 71} 141

\bibitem{Butter} Hofstadter D R, 1976 {\it Phys. Rev. B} {\bf 14} 2239

\bibitem{Guinea} Gonzales J, Guinea F and Vozmediano M A H, 1993 {\it Phys. Rev. Lett.} {\bf 69} 172
\nonum
Gonzales J, Guinea F and Vozmediano M A H, 1993 {\it Nucl. Phys. B} {\bf 406} 771

\bibitem{Sondhi} Castelnovo C, Moessner R and Sondhi S L, 2008 {\it Nature} {\bf 451} 42

\bibitem{Haldane} Haldane F D M, 1983 {\it Phys. Rev. Lett.} {\bf 51} 605

\bibitem{AHK} Avishai Y, Hatsugai Y and Kohmoto M, 1995 {\it Phys. Rev. B} {\bf 51} 13419

\bibitem{Dirac} Dirac P A M, 1931 {\it Proc. R. Soc. London A} {\bf 133} 60

\bibitem{Yang} Wu T T and Yang C N, 1975 {\it Phys. Rev. D} {\bf 12} 3845

\bibitem{wilson} Wilson R J, 1979 {\it Introduction to Graph Theory} 2nd ed
(London: Longman)

\bibitem{stree} M\"uller V F and R\"uhl W, 1984 {\it Nucl. Phys. B} {\bf 230} 49

\bibitem{sm} Sadoc J F and Mosseri R, 1999 {\it Geometrical Frustration} Monographs and Texts in Statistical Physics (New York: Cambridge University Press)

\bibitem{hase} Hasegawa Y, Lederer P, Rice T M and Wiegmann P B, 1989 {\it Phys. Rev. Lett.} {\bf 63} 907

\bibitem{lengths} Hedberg K, Hedberg L, Bethune D S, Brown C A, Dorn H C, Johnson R D and Devries M, 1991 {\it Science} {\bf 254} 410

\bibitem{manou} Manousakis E, 1991 {\it Phys. Rev. B} {\bf 44} 10991

\bibitem{fibospectrum} Albuquerque E L and Cottam M G, 2003 {\it Phys. Rep} {\bf 376} 225

\bibitem{graphene_review} Castro-Neto A H, Guinea F, Peres N M R, Novoselov K S and Geim A K, 2008 {\it Rev. Mod. Phys.} at press Preprint arXiv:0709.1163

\bibitem{LLG} McClure J W, 1956 {\it Phys. Rev.} {\bf 104} 666
\nonum Semenoff G W, 1984 {\it Phys. Rev. Lett.} {\bf 53} 2449
\nonum Haldane F D M, 1988 {\it Phys. Rev. Lett.} {\bf 61} 2015

\bibitem{kk} Korn G A and Korn T M, 1968 {\it Mathematical Handbook for Scientists and Engineers} (New York: McGraw-Hill)

\bibitem{edm} Ito K ed, 1980 {\it The Encyclopedic Dictionary of Mathematics} 2nd ed (Cambridge, MA: MIT Press)

\end{thebibliography}
\end{document}